\documentclass{eas}
\usepackage{graphicx}
\begin{document}
%
\title{Modelling the light-curves of objects tidally disrupted by a black hole} 
\author{Uro\v s Kosti\' c}\address{Department of Physics, Faculty of Mathematics and Physics, University of Ljubljana, Jadranska 19, SI-1000 Ljubljana, Slovenia}
\author{Andreja Gomboc}\sameaddress{1}
\author{Andrej \v Cade\v z}\sameaddress{1}
\author{Massimo Calvani}\address{INAF, Astronomical Observatory of Padova, Vicolo dell'Osservatorio 5, I-35122 Padova, Italy}
\begin{abstract}
Tidal disruption by massive black holes is a phenomenon, during which a large part of gravitational energy can be released on a very short time-scale. The time-scales and energies involved during X-ray and IR flares observed in Galactic centre suggest that they may be related to tidal disruption events. Furthermore, aftermath of a tidal disruption of a star by super-massive black hole has been observed in some galaxies, e.g. RX J1242.6-1119A. All these discoveries increased the demand for tools for tidal disruption study in curved space-time. Here we summarise our study of general relativistic effects on tidal deformation of stars and compact objects.
\end{abstract}
\maketitle
\section{Model of tidal deformation}
The gravitational field of a black hole can induce strong tidal effects on objects which find themselves below their Roche radius. It has been shown (Carter \& Luminet \cite{carter1}, \cite{carter2}) that inside the Roche radius the gravity of the black hole dominates over self gravity and hydrodynamics. If the accretion event occurs from a highly eccentric orbit, tidal energy released during a close encounter may yield up to a few percent of $mc^2$ (Gomboc \& \v Cade\v z \cite{gomboc}). In such cases the result is total tidal disruption of the object which is deformed into long thin spiral thread.
\subsection{Light curve modelling}
We consider an object as a sphere approaching a black hole. After Roche radius crossing pressure forces become negligible with respect to inertial forces and the particles that constitute the object start to fall freely in the gravitational field of the black hole. We use very efficient ray-tracing from points on the object's surface to the distant observer to determine its appearance and light-curve. To determine geodesics which connect two points in space-time, we calculate the exact analytical solutions of relativistic orbit equations, including the calculation of the time of flight. Ray tracing procedures are described in details in \v Cade\v z \& Kosti\' c (\cite{cadez1}) and are freely available upon request from authors. To calculate the light-curves we take into account primary and secondary image only, since contribution of higher order images is negligible. We also take into account relativistic effects, such as gravitational redshift, Doppler blue/redshift and aberration of light. The images are calculated at equal observer's times for primary and secondary photons.

\subsection{Results on tidal disruption of a star}

Tidal disruption of a gaseous star has been studied by a number of authors (Carter \& Luminet \cite{carter1}, \cite{carter2}; Laguna et al. \cite{laguna}; Luminet \& Marck \cite{luminet}) with emphasis on stellar structure during the encounter. How such an event would be seen, i.e. luminosity variations occurring to the star in the vicinity of the black hole, is discussed in Gomboc \& \v Cade\v z (\cite{gomboc}).
\subsection{Results on tidal disruption of asteroid-like objects}
We also study tidal effects of a black hole on small compact objects (e.g. asteroids). It has been shown (\v Cade\v z et al. \cite{cadez2}) that solid objects can melt due to work done by tidal forces. Once melted, the object is prone to tidal deformations. Results show  that relativistic tidal effects depend predominantly on the ratio $\zeta=(E/mc^2-V_{\rm min})/(V_{\rm max}-V_{\rm min})$. Here E is the energy of the object. $V_{\rm min}$ and $V_{\rm max}$ are the minimum and the maximum of the relativistic effective potential, respectively.  Tidal deformations grow exponentially with time for infall of a deformable object on a critical orbit, i.e. orbit with $\zeta\sim 1$.

The light-curves show a slow component which is due to increasing tidal deformations, and a fast component which is represented by sharp peaks due to relativistic effects. Both components depend also on the orientation of the orbit with respect to the observer. More details in Kosti\' c (\cite{kostic}).

\end{document}